\documentclass[aps,twocolumn,showpacs,floatfix,amsmath,amsfonts]{revtex4}
\usepackage{graphicx}

\newcommand{\bra}[1]{\langle #1 | \,}
\newcommand{\ket}[1]{\, | #1 \rangle}

\newcommand{\expv}[1]{\langle #1 \rangle}
\newcommand{\ddt}{\frac{\partial}{\partial t}}
\newcommand{\ddz}{\frac{\partial}{\partial z}}
\newcommand{\lra}{\leftrightarrow}
\newcommand{\la}{\lambda}
\newcommand{\La}{\Lambda}
\newcommand{\om}{\omega}
\newcommand{\Om}{\Omega}
\newcommand{\ga}{\gamma}
\newcommand{\Ga}{\Gamma}
\newcommand{\de}{\delta}
\newcommand{\De}{\Delta}
\newcommand{\ka}{\kappa}
\newcommand{\sih}{\hat{\sigma}}

\newcommand{\Eh}{\hat{\cal E}}
\newcommand{\Psih}{\hat{\Psi}}
\newcommand{\psih}{\hat{\psi}}
\newcommand{\Fh}{\hat{F}}
\newcommand{\Ih}{\hat{I}}
\newcommand{\Fch}{\hat{\cal F}}
\newcommand{\zp}{z^{\prime}}

\newcommand{\eps}{\epsilon}

\begin{document}

\title{Towards deterministic optical quantum computation \\
with coherently driven atomic ensembles}

\author{David Petrosyan}
\email[E-mail: ]{dap@iesl.forth.gr}
\affiliation{Institute of Electronic Structure \& Laser, 
FORTH, Heraklion 71110, Crete, Greece}

\date{\today}

\begin{abstract}
Scalable and efficient quantum computation with photonic qubits 
requires (i) deterministic sources of single-photons, (ii)
giant nonlinearities capable of entangling pairs of photons,
and (iii) reliable single-photon detectors. In addition, an
optical quantum computer would need a robust reversible photon
storage devise. Here we discuss several related techniques, 
based on the coherent manipulation of atomic ensembles 
in the regime of electromagnetically induced transparency, that
are capable of implementing all of the above prerequisites for 
deterministic optical quantum computation with single photons.
\end{abstract}


\maketitle

\section{Introduction}

The field of quantum information, which extends and generalizes the 
classical information theory, is currently attracting enormous interest 
in view of its fundamental nature and its potentially revolutionary 
applications to computation and secure communication \cite{QCQI,QCQIrev}. 
Among the various schemes for physical implementation of quantum 
computation \cite{solst,DP-GK,iontr,BCJD,phphcav,linopt}, those based 
on photons \cite{phphcav,linopt} have the advantage of using very robust 
and versatile carriers of quantum information. However, the absence of 
practical single-photon sources and the weakness of optical nonlinearities 
in conventional media \cite{Boyd} are the major obstacles for the realization 
of efficient all-optical quantum computation. To circumvent these difficulties,
it has been proposed to use linear optical elements, such as beam-splitters 
and phase-shifters, in conjunction with parametric dawn-converters and 
single-photon detectors, for achieving probabilistic quantum logic with 
photons conditioned on the successful outcome of a measurement performed 
on auxiliary photons \cite{linopt}. Yet, an efficient and scalable device 
for quantum information processing with photons would ideally require 
deterministic sources of single photons, strong nonlinear photon-photon 
interaction and reliable single-photon detectors. In addition, a 
versatile optical quantum computer would need a robust reversible 
memory devise. 

In this paper we discuss several related techniques which can be used
to implement all of the above prerequisites for deterministic optical 
quantum computation. The schemes discussed below are based on the coherent 
manipulation of atomic ensembles in the regime of electromagnetically 
induced transparency (EIT) \cite{eit_rev,ScZub,eitbw,lukin},
which is a quantum interference effect that results in a dramatic 
reduction of the group velocity of weak propagating field accompanied 
by vanishing absorption \cite{vred}. As the quantum interference is 
usually very sensitive to the system parameters, various schemes 
exhibiting EIT have recently received considerable attention due to 
their potential for greatly enhancing nonlinear optical effects. 
Some of the most representative examples include slow-light enhancement 
of acusto-optical interactions in doped fibers \cite{acopt}, trapping 
light in optically dense atomic and doped solid-state media by coherently 
converting photonic excitation into spin excitation \cite{fllk,v0exp,hemmer} 
or by creating photonic band gap via periodic modulation of the EIT 
resonance \cite{lukin-pbg}, and nonlinear photon-photon coupling using 
N-shaped configuration of atomic levels \cite{imam,haryam,harhau,lukimam}.

Below, we will focus on the optical implementation of quantum computation 
with qubit basis states represented by two orthogonal polarization states 
of single photons, as opposed to an alternative approach, wherein 
nearly-orthogonal weak coherent states of optical fields are used 
\cite{QI-ContVar,QC-CohSt}. The chief motivation for this is 
that single-photon pulses provide a natural choice for qubits employed 
in quantum computation and quantum communication protocols \cite{QCQI,QCQIrev},
and facilitate the convenience and intuitiveness in the description of 
their dynamics in quantum information processing networks. 

In section~\ref{sec:oqc} we outline the envisioned setup of an optical 
quantum computer and discuss the physical implementations of the required 
single- and two-qubit logic operations. Section~\ref{sec:eit} gives a 
concise introduction of the EIT in optically dense atomic media, which 
is necessary for understanding the principles of operation of photonic 
memory devise of section~\ref{sec:mem}, deterministic single photon 
sources discussed in section~\ref{sec:sphs}, giant cross-phase modulation 
of section~\ref{sec:xpm}, and reliable single-photon detection presented 
in section~\ref{sec:sphd}. The conclusions are summarized in 
section~\ref{sec:sum}.

\section{Optical quantum computer}
\label{sec:oqc}

\begin{figure*}[t]
\centerline{\includegraphics[width=15cm]{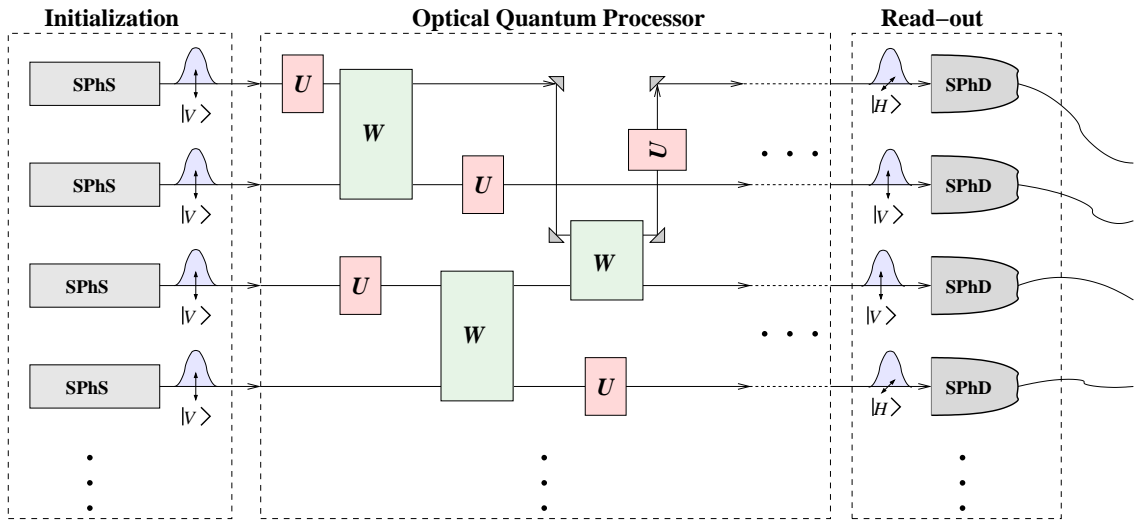}}
\caption{Schematic representation of the quantum computer with 
single-photon qubits. The operation of the computer consists of the
following principal steps: {\it Qubit initialization} is realized by 
deterministic single-photon sources (SPhS). {\it Information processing}
is implemented by the quantum processor with single-qubit $U$ and 
two-qubit $W$ logic gates. {\it Read-out} of the result of computation 
is accomplished by efficient single-photon detectors (SPhD).}
\label{fig:optQC}
\end{figure*}

Quantum computer is an envisaged physical device for processing the
information encoded in a collection of two-level quantum-mechanical 
systems -- qubits -- quantum analogs of classical bits. 
Such a computer would typically be composed of (a)~quantum register 
containing a number of qubits, whose computational basis states 
are labeled as $\ket{0}$ and $\ket{1}$; (b)~one- and two-qubit 
(and possibly multi-qubit) logic gates -- unitary operations applied 
to the register according to the particular algorithm; and (c)~measuring 
apparatus applied to the desired qubits at the end of (and, possibly, during) 
the program execution, which project the qubit state onto the computational 
basis $\{\ket{0}, \ket{1} \}$. Operation of the quantum computer may 
formally be divided into the following principal steps. 
{\it Initialization}: Preparation of all qubits of the register 
in a well-defined initial state, such as, e.g., $\ket{0 \ldots 000}$. 
{\it Input}: Loading the input data using the logic gates.
{\it Computation}: The desired unitary transformation of the register. 
Any multiqubit unitary transformation can be decomposed into a sequence of 
single-qubit rotations and two- (or more) qubit conditional operations,
which thus constitute the universal set of quantum logic gates.
{\it Output}: Projective measurement of the final state of the 
register in the computational basis. The reliable measurement scheme
would need to have the fidelity more than $1/2$, but ideally as close 
to 1 as possible.

A schematic representation of an optical quantum computer is shown 
in figure~\ref{fig:optQC}. In the initialization section of the computer, 
deterministic sources of single photons generate single-photon pulses 
with precise timing and well-defined polarization and pulse-shapes
(see section~\ref{sec:sphs}). A collection of such photons constitutes 
the quantum register. The qubit basis states $\{\ket{0}, \ket{1} \}$ of 
the register are represented by the vertical $\ket{V} \equiv \ket{0}$ 
and horizontal $\ket{H} \equiv \ket{1}$ polarization states of the 
photons. The preparation of an initial state of the register and the 
execution of the program according to the desired quantum algorithm 
is implemented by the quantum processor. This amounts to the application 
of certain sequence of single-qubit $U$ and two-qubit $W$ unitary operations,
whose physical realization is described below. Finally, the result of  
computation is read-out by a collection of efficient polarization-sensitive 
photon detectors (see section~\ref{sec:sphd}).

For the photon-polarization qubit $\ket{\psi}=\alpha \ket{V} + \beta \ket{H}$,
the universal set of quantum gates can be constructed from arbitrary 
single-qubit rotation operations $U$ and a two-qubit conditional operation 
$W$, such as the controlled-{\sc not} ({\sc cnot}) operation 
$\ket{a} \ket{b} \to \ket{a} \ket{a \oplus b}$ ($a,b \in \{ 0,1 \}$) 
or controlled-phase or $Z$ ({\sc cphase} or {\sc cz}) operation 
$\ket{a} \ket{b} \to (-1)^{ab} \ket{a} \ket{b}$.
In turn, any single-qubit unitary operation $U$ can be decomposed
into a product of rotation $R(\theta)$ and phase-shift $T(\phi)$ 
operations
\[
R(\theta) =  \left[ \begin{array}{cc}
\cos \theta & - \sin \theta \\ 
\sin \theta & \cos \theta 
\end{array} \right] , \;\;\;
T(\phi) =  \left[ \begin{array}{cc}
1 & 0 \\ 
0 & e^{i \phi} 
\end{array} \right] ,  
\]
and an overall phase shift $e^{i \varphi}$. As an example, the Pauli
$X$, $Y$, $Z$ and Hadamard $H$ transformations can be represented as
$X = R(\pi/2) T(\pi)$, $Y = e^{i \pi /2} R(\pi/2)$, $Z = T(\pi)$, 
$H = R(\pi/4) T(\pi)$. 

\begin{figure}[t]
\centerline{\includegraphics[width=8.5cm]{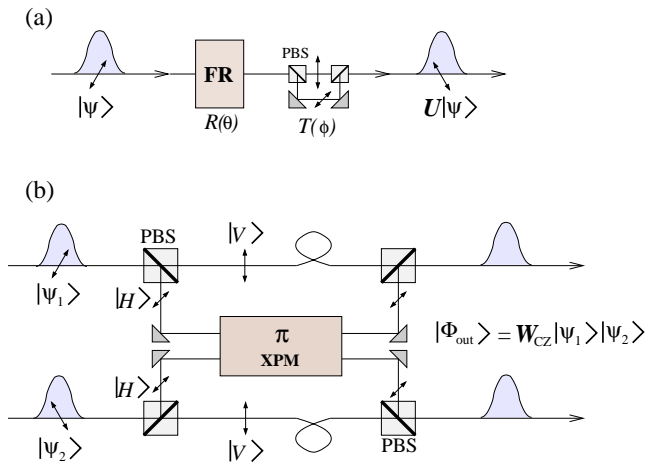}}
\caption{Proposed physical implementation of quantum logic gates.
(a) Single-qubit logic gates $U$ are implemented with a sequence
of two linear-optics operations: $R(\theta)$ -- Faraday rotation (FR) 
of photon polarization by angle $\theta$ about the propagation direction;
$T(\phi)$ -- relative phase-shift $\phi$ of the photon's $\ket{V}$ and 
$\ket{H}$ polarized components due to their optical paths difference.
(b) Two-qubit {\sc cz} (or {\sc cphase}) gate $W_{\textsc{cz}}$ is
realized using polarizing beam-splitters (PBS) and $\pi$ cross-phase
modulation studied in section~\ref{sec:xpm}.}
\label{fig:UWrztn}
\end{figure}

As shown in figure~\ref{fig:UWrztn}(a), for the photon-polarization qubit 
$\ket{\psi}$ the $R(\theta)$ and $T(\phi)$ operations are implemented, 
respectively, by the rotation of photon polarization by angle 
$\theta$ about the propagation direction, and relative phase-shift 
$\phi$ of the $\ket{V}$ and $\ket{H}$ polarized components of the photon. 
Both operations are easy to implement with the standard linear optical 
elements, Faraday rotators, polarizing beam-splitters or phase-retardation 
(birefringent) waveplates. A possible realization of the {\sc cphase} 
two-qubit entangling operation is shown in figure~\ref{fig:UWrztn}(b).
There, after passing through a polarizing beam-splitter, the vertically 
polarized component of each photon is transmitted, while the horizontally 
polarized component is directed into the active medium, wherein
the two-photon state $\ket{\Phi_{\rm in}} = \ket{H_1 \, H_2}$ acquires 
the conditional phase-shift $\pi$, as discussed in section~\ref{sec:xpm}. 
At the output, each photon is recombined with its vertically polarized 
component on another polarizing beam-splitter, where the complete temporal 
overlap of the vertically and horizontally polarized components of each 
photon is achieved by delaying the $\ket{V}$ wavepacket in a fiber loop 
or sending it though a EIT vapor cell in which the pulse propagates with
a reduced group velocity (see section~\ref{sec:eit}).  

The remainder of this paper is devoted to the physical realizations of
the constituent parts of the optical quantum computer described above.

\section{Electromagnetically induced transparency}
\label{sec:eit}

\begin{figure}[t]
\centerline{\includegraphics[width=8.5cm]{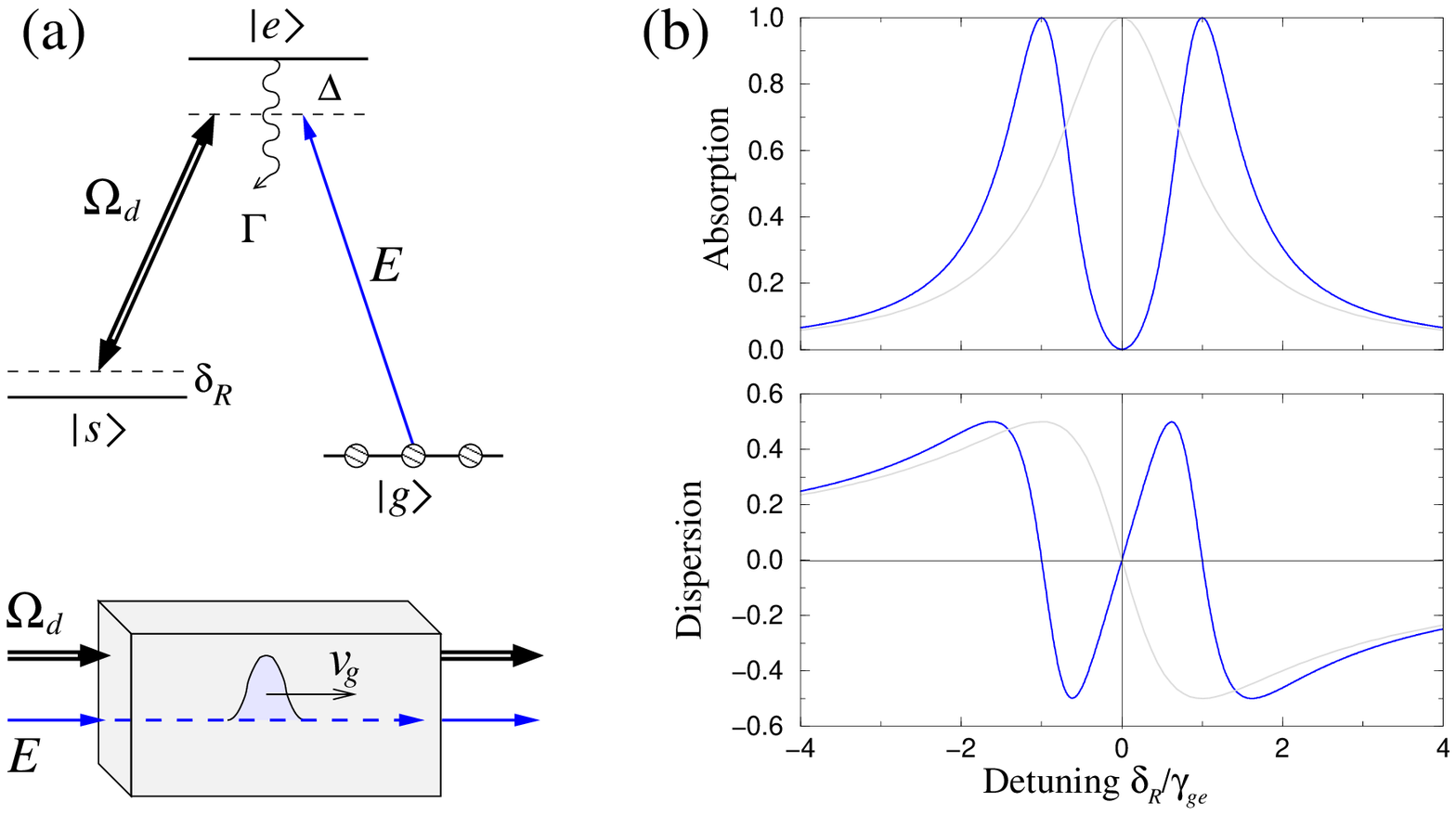}}
\caption{Electromagnetically induced transparency in atomic medium.
(a)~Level scheme of three-level $\La$-atoms interacting with a cw
driving field with Rabi frequency $\Om_d$ on the transition 
$\ket{s} \lra \ket{e}$ and a weak pulsed $E$ field acting on the 
transition $\ket{g} \lra \ket{e}$. The lower states $\ket{g}$ and
$\ket{s}$ are long-lived (metastable), while the excited state 
$\ket{e}$ decays fast with the rate $\Ga$.
(b)~Absorption and dispersion spectra ($\de_R = \De$) of the atomic 
medium for the $E$ field in units of resonant absorption coefficient 
$\ka_0$, for $\Om_d/\ga_{ge} = 1$ and $\ga_R/\ga_{ge} = 10^{-3}$. 
The light-gray curves correspond to the case of $\Om_d = 0$ (two-level atom).}
\label{fig:eit3la}
\end{figure}

The propagation of a weak probe field $E e^{i(kz -\om t)}$ with 
carrier frequency $\om$ and wave vector $k=\om/c$ in a near-resonant medium 
can be characterized by the linear susceptibility $\chi(\om)$, whose 
real and imaginary parts describe, respectively, the dispersive and 
absorptive properties of the medium: $E(z) = E(0) e^{-\kappa z} e^{i \phi(z)}$,
where $\kappa = k/2 \, {\rm Im} \chi(\om)$ is the linear
absorption coefficient, and $\phi(z) = k/2 \, {\rm Re} \chi(\om) z$
is the phase-shift. In the case of light interaction with two-level atoms 
on the transition $\ket{g} \to \ket{e}$, the familiar Lorentzian absorption 
spectrum leads to the strong attenuation of the resonant field 
($\De \equiv \om -\om_{eg} =0$) in the optically dense medium according to 
$E(z) = E(0) e^{-\ka_0 z}$, where the resonant absorption coefficient 
$\ka_0 = \sigma_0 \rho$ is given by the product of atomic density $\rho$ 
and absorption cross-section  
$\sigma_0 = \wp_{ge}^2 \om / (2 \hbar c \eps_0 \ga_{ge})$, 
$\wp_{ge} = \bra{g} \textbf{d} \ket{e}$ being the dipole matrix element 
for the transition $\ket{g} \to \ket{e}$ and
$\ga_{ge}(\geq \Ga/2)$ the corresponding coherence relaxation rate.
When, however, the excited state $\ket{e}$ having decay rate $\Ga$ 
is coupled by a strong driving field with Rabi frequency $\Om_d$ and detuning 
$\De_d = \om_d - \om_{es}$ to a third metastable state $\ket{s}$, the 
situation changes dramatically (figure~\ref{fig:eit3la}(a)). Assuming
all the atoms initially reside in state $\ket{g}$, the complex 
susceptibility now takes the form
\begin{equation}
\chi(\om) = \frac{2 \ka_0}{k}\frac{i \ga_{ge}}{\ga_{ge} -i \De + 
|\Om_d|^2 (\ga_R - i \de_R)^{-1}} , \label{susc}
\end{equation}
where $\de_R = \De - \De_d = \om - \om_d -\om_{sg}$ is the two-photon Raman
detuning and $\ga_R$ the Raman coherence (spin) relaxation rate. Obviously, 
in the limit of $\Om_d \to 0$, the susceptibility (\ref{susc}) reduces to 
that for the two-level atom. The absorption and dispersion spectra 
corresponding to the susceptibility of equation~(\ref{susc}) are shown 
in figure~\ref{fig:eit3la}(b) for the case of $\Om_d =\ga_{ge}$ and 
$\De_d = 0$, i.e. $\de_R = \De$. As seen, the interaction with the driving 
field results in the Autler-Towns splitting of the absorption spectrum into 
two peaks separated by $2 \Om_d$, while at the line center the medium becomes 
transparent to the resonant field, provided $\ga_R \ll |\Om_d|^2/\ga_{ge}$. 
This effect is called electromagnetically induced transparency (EIT) 
\cite{eit_rev,ScZub}. At the exit from the optically dense medium of length 
$L$ (optical depth $2 \ka_0 L >1$), the intensity transmission coefficient 
is given by $T(\om) = \exp[- k \, {\rm Im} \chi(\om) L]$. To determine the 
width of the transparency window $\de \om_{\rm tw}$, one makes a power series 
expansion of ${\rm Im} \chi(\om)$ in the vicinity of maximum transmission 
$\de_R = 0$, obtaining \cite{eitbw,lukin}
\begin{equation}
T(\om) \simeq \exp(-\de_R^2/\de \om^2_{\rm tw}) , \;\;\; 
\de \om_{\rm tw} = \frac{|\Om_d|^2}{\ga_{ge}\sqrt{2 \ka_0 L}} , \label{eit_bw}
\end{equation}
where the usual EIT conditions, 
$\De_d \ga_R,\De_d^2 \ga_R/\ga_{ge} \ll |\Om_d|^2$, are assumed satisfied.
Equation~(\ref{eit_bw}) implies that for the absorption-free propagation,
the bandwidth $\de \om$ of near-resonant probe field should be within 
the transparency window, $\de \om < \de \om_{\rm tw}$. Alternatively, the 
temporal width $T$ of a Fourier-limited probe pulse should satisfy 
$T \gtrsim \de \om_{\rm tw}^{-1}$. 

Considering next the dispersive properties of EIT, in  
figure~\ref{fig:eit3la}(b) one sees that the dispersion exhibits a steep 
and approximately linear slope in the vicinity of absorption minimum
$\de_R = 0$. Therefore, a probe field slightly detuned from the resonance 
by $\de_R < \de \om_{\rm tw}$, during the propagation would acquire a large 
phase-shift $\phi(L) \simeq \ka_0 L \ga_{ge} \de_R/|\Om_d|^2$ while 
suffering only little absorption, as per equation~(\ref{eit_bw}).
At the same time, a near-resonant probe pulse $E(z,t)$ propagates 
through the medium with greatly reduced group velocity 
\begin{equation}
v_g = \frac{c}{1 + 
c \frac{\partial}{\partial \om} [\frac{k}{2} {\rm Re} \chi(\om)]} =
\frac{c}{1+c \frac{\ka_0 \ga_{ge}}{|\Om_d|^2}} 
\simeq \frac{|\Om_d|^2}{\ka_0 \ga_{ge}} \ll c , \label{v_gr} 
\end{equation}
while upon entering the medium, its spacial envelope is compressed 
by a factor of $v_g/c\ll 1$.

So far, we have outlined the absorptive and dispersive properties of
the stationary EIT without elaborating much on the underlying physical
mechanism. As elucidated below, EIT is based on the phenomenon of coherent 
population trapping \cite{eit_rev,ScZub}, in which the application of two
coherent fields to a three-level $\La$ system of figure~\ref{fig:eit3la}(a) 
creates the so-called ``dark state'', which is stable against absorption 
of both fields. Since we are interested in quantum information processing
with photons, the probe field has to be treated quantum mechanically. It
is expressed through the traveling-wave (multimode) electric field operator 
$\Eh(z,t) = \sum_q a^q(t) e^{iqz}$, where $a^q$ is the bosonic annihilation 
operator for the field mode with the wave-vector $k+q$. The classical driving 
field with Rabi frequency $\Om_d$ is assumed spatially uniform. In the 
frame rotating with the probe and driving field frequencies, the interaction 
Hamiltonian has the following form:
\begin{eqnarray}
H &=& \hbar \sum_{j=1}^N [\De \sih_{ee}^j + \de_R \sih_{ss}^j 
-g \Eh(z_j) e^{ikz_j} \sih_{eg}^j 
\nonumber \\ & & \;\;\;\;\;\;\;\;\;
- \Om_d(t) e^{ik_d z_j}\sih_{es}^j + {\rm H. c.}] , \label{ham}
\end{eqnarray}
where $N = \rho V$ is the total number of atoms in the quantization 
volume $V = A L$ ($A$ being the cross-sectional area of the probe field), 
$\sih_{\mu\nu}^j = \ket{\mu_j}\bra{\nu_j}$ is the transition operator of 
the $j$th atom at position $z_j$, $k_d$ is the projection of the driving 
field wavevector onto the $\textbf{e}_z$ direction, and  
$g = \frac{\wp_{ge}}{\hbar} \sqrt{\frac{\hbar \om}{2 \eps_0 V}}$ is the 
atom-field coupling constant. For $\de_R = 0$, the Hamiltonian~(\ref{ham}) 
has a family of dark eigenstates $\ket{D^q_n}$ with zero eigenvalue 
$H \ket{D^q_n} = 0$, which are decoupled from the rapidly decaying excited 
state $\ket{e}$:  
\begin{equation}
\ket{D^q_n} = \sum_{m=0}^n 
\left(\begin{array}{c} n \\ m \end{array} \right)^{\frac{1}{2}} 
(- \sin \theta)^{m} (\cos \theta)^{n-m} \ket{(n-m)^q} \ket{s^{(m)}} .  
\label{gdarkst}
\end{equation}
Here the mixing angle $\theta(t)$ is defined via 
\[
\tan^2 \theta(t) = \frac{g^2 N}{|\Om_d(t)|^2} \, ,
\]
$\ket{n^q}$ denotes the state of the quantum field with $n$ photons
in mode $q$, and $\ket{s^{(m)}}$ is a symmetric Dicke-like state of the 
atomic ensemble with $m$ Raman (spin) excitations, i.e., atoms in state 
$\ket{s}$, defined as 
\begin{eqnarray*}
\ket{s^{(0)}} & \equiv & \ket{g_1,g_2,\ldots,g_N} , \\
\ket{s^{(1)}} & \equiv & \frac{1}{\sqrt{N}} \sum_{j=1}^N 
e^{i(k+q-k_d) z_j} \ket{g_1,\ldots,s_j, \ldots,g_N} , \\
\ket{s^{(2)}} & \equiv & \frac{1}{\sqrt{2N(N-1)}} 
\sum_{i\neq j=1 }^N e^{i(k+q-k_d) (z_i+z_j)} 
\\ & & \;\;\;\;\;\;\;\;\;\;\;\;\;\;\;\;\times
\ket{g_1,\ldots,s_i,\ldots,s_j, \ldots,g_N},
\end{eqnarray*}
etc. When $\theta = 0$ ($|\Om_d|^2 \gg g^2 N$), the dark state 
(\ref{gdarkst}) is comprised of purely photonic excitation, 
$\ket{D^q_n} = \ket{n^q} \ket{s^{(0)}}$, while in the opposite limit of 
$\theta = \pi/2$ ($|\Om_d|^2 \ll g^2 N$), it coincides with the collective
atomic excitation $\ket{D^q_n} = (-1)^n \ket{0^q} \ket{s^{(n)}}$. For 
intermediate values of mixing angle $0 < \theta <\pi/2$, the dark state 
represents a coherent superposition of photonic and atomic Raman 
excitations \cite{fllk,LYF}. Below  will be concerned with the case of 
single-photon probe field, for which the dark state takes a particularly 
simple form
\begin{equation}
\ket{D^q_1} = \cos \theta \ket{1^q,s^{(0)}} 
- \sin \theta \ket{0^q,s^{(1)}} . \label{sphdarkst}
\end{equation}

Consider now the dynamic evolution of the field and atomic operators.
In the slowly varying envelope approximation, the propagation equation
for the quantum field has the form
\begin{equation}
\left(\ddt + c \ddz \right) \Eh(z,t) = i g N \sih_{ge},\label{Eprop} 
\end{equation}
where $\sih_{\mu \nu}(z,t)=\frac{1}{N_z} \sum_{j=1}^{N_z} \sih_{\mu \nu}^j$ 
is the collective atomic operator averaged over small but macroscopic volume 
containing many atoms $N_z = (N/L) dz \gg 1$ around position $z$. The
evolution of the atomic operators is governed by a set of Heisenberg-Langevin 
equations \cite{fllk}, which are treated perturbatively in the small parameter 
$g \Eh/\Om_d$ and in the adiabatic approximation for both fields,
\begin{subequations}
\label{ss}
\begin{eqnarray}
\sih_{ge} &=& -\frac{i}{\Om_d^{*}} \left[ \ddt - i \de_R + \ga_R \right] 
\sih_{gs} + \frac{i}{\Om_d^{*}} \Fh_{gs}, \label{ssge} \\
\sih_{gs} &=&  - \frac{g \Eh}{\Om_d} 
\left[ 1+ \frac{\de_R (\De +i\ga_{ge})}{|\Om_d|^2} \right] 
+ \frac{i}{\Om_d} \Fh_{ge} , \label{ssgs} 
\end{eqnarray}
\end{subequations}
where $\Fh_{\mu \nu}$ are $\de$-correlated noise operators associated with 
the atomic relaxation. When the driving field is constant in time and 
$\de_R \De \ll |\Om_d|^2$, equations~(\ref{Eprop}-\ref{ss}) yield
\begin{equation}
\left(\ddz + \frac{1}{v_g} \ddt \right) \Eh 
= - \ka \Eh +i s \de_R \Eh + \Fch ,\label{EprOmdc} 
\end{equation}
where $v_g = c \cos^2 \theta$ is the group velocity of (\ref{v_gr}), 
while $\ka =\tan^2 \theta /c (\ga_R + \ga_{ge} \de_R^2/|\Om_d|^2)$ and 
$s = \tan^2 \theta /c$ are, respectively, the linear absorption and 
phase-modulation coefficients. The solution of equation~(\ref{EprOmdc}) 
can be expressed in terms of the retarded time $\tau = t -z/v_g$ as
\begin{equation}
\Eh(z,t) = \Eh(0,\tau) \exp \left[ -\ka z + i \phi(z) \right] + 
\Fch_{\mathcal{E}} , \label{solEprOmdc}
\end{equation}
with $\phi(z) = s\de_R z$ (the noise operator $\Fch_{\mathcal{E}}$ 
ensures the conservation of field commutators \cite{DPYuM}).
We will be interested in the input states corresponding to single-photon 
wavepackets $\ket{1} = \sum_{q} \xi^q \ket{1^q}$ 
($\ket{1^q} = a^{q \dagger} \ket{0}$), where
the Fourier amplitudes $\xi^q$, normalized as $\sum_{q} |\xi^q|^2 =1$, 
define the spatial envelope $f(z)$ of the probe pulse that initially 
(at $t=0$) is localized around $z=0$,
\[
\bra{0} \Eh(z,0) \ket{1} = \sum_{q} \xi^q e^{iqz} = f(z).
\]
In free space, $\Eh(z,t) =\Eh(0,\tau)$ with $\tau = t -z/c$, and we have 
$\bra{0} \Eh(z,t) \ket{1} = f(z-ct)$. Upon propagating through the EIT 
medium, using equation~(\ref{solEprOmdc}) and neglecting the (small) 
absorption, for the expectation value of the probe field intensity 
$\expv{\Ih(z,t)} = \bra{1}\Eh^{\dagger}(z,t) \Eh(z,t) \ket{1}$ one has
\begin{equation}
\expv{\Ih(z,t)} = |f(-c \tau)|^2 = |f(z c /v_g - ct)|^2 , 
\label{res-evI}
\end{equation}
where $\tau = t - z/v_g$ for $0 \leq z < L$. This equation indicates that
at the entrance to the medium, as the group velocity of the pulse is slowed 
down to $v_g$, its spatial envelope is compressed by a factor of 
$v_g/c \ll 1$. Outside the medium, at $z \geq  L$ and accordingly 
$\tau = t - L/v_g -(z-L)/c$, one has 
$\expv{\Ih(z,t)} = |f(z + L(c/v_g -1) -ct)|^2$, which shows that
the propagation velocity and the pulse envelope are restored to their
free-space values. 

Consider finally the case of the exact two-photon Raman resonance 
$\de_R = 0$ and time-dependent driving field $\Om_d(t)$. To solve the coupled 
set of equations~(\ref{Eprop}-\ref{ss}), one introduces a polariton operator 
\cite{fllk}
\begin{equation}
\Psih(z,t) = \cos \theta(t) \Eh(z,t) - \sin \theta(t) \sqrt{N} \sih_{gs} , 
\label{polar}
\end{equation}
whose photonic and atomic components are determined by the mixing
angle $\theta$, $\Eh = \cos \theta \Psih$ and 
$\sih_{gs} = \sin \theta \Psih /\sqrt{N}$. Taking the plain-wave decomposition 
of the polariton operator $\Psih(z,t) = \sum_q \psih^q(t) e^{iqz}$, one can 
show that in the weak-field limit the mode operators $\psih^q$ obey the 
bosonic commutation relations 
$[\psih^q,\psih^{q^\prime \dagger}] = \de_{qq^{\prime}}$ \cite{fllk}.
Moreover, by acting $n$ times with operator $\psih^{q \dagger}$ onto the 
state $\ket{0^q} \ket{s^{(0)}}$ one creates the dark state of (\ref{gdarkst}), 
\[
\ket{D^q_n} = \frac{1}{\sqrt{n!}} 
(\psih^{q \dagger})^n \ket{0^q} \ket{s^{(0)}}.
\]
Therefore the operator $\Psih$ had been called dark-state polariton 
\cite{fllk}. It is easy to verify that upon neglecting the absorption, 
the equation of motion for $\Psih(z,t)$ takes a particularly simple form 
\begin{equation}
\left(\ddt + v_g(t) \ddz \right) \Psih(z,t) = 0 . 
\label{polareqmot}
\end{equation}
Its solution is given by 
\begin{equation}
\Psih(z,t) = \Psih\left( z- \int_0^t v_g(t^{\prime}) d t^{\prime}, 0 \right), 
\label{polsol}
\end{equation}
which describes a state- and shape-preserving pulse propagation with  
time-dependent group velocity $v_g (t) = c \cos^2 \theta(t)$. Thus, once
the pulse has fully accommodated in the medium, one can stop it by 
adiabatically rotating the mixing angle from its initial value
$0 \leq \theta < \pi/2$ to $\theta = \pi/2$, which amounts to switching
off the driving field $\Om_d$. As a result, the state of the photonic
component of the pulse is coherently mapped onto the collective atomic 
state according to (\ref{gdarkst}) or (\ref{sphdarkst}), the latter
applies to a single-photon input pulse. In order to accommodate the pulse 
in the medium with negligible losses, its duration should exceed the 
inverse of the initial EIT bandwidth, while at the entrance its length 
should be compressed to the length of the medium,
$\de \om_{\rm tw}^{-1} v_g\ll T v_g < L$. These two conditions yield
$(2 \ka_0 L)^{-1/2} \ll T v_g/L < 1$, which requires media with large
optical depth $2 \ka_0 L \gg 1$. Note finally, that although the collective
state $\ket{s^{(1)}}$ is an entangled state of $N$ atoms, it decoheres with
essentially single-atom rate and is quite stable against one (or few) 
atom losses \cite{MFCM}. Therefore the coherent trapping time is limited 
mainly by the life-time of Raman coherence $\ga_R^{-1}$.

\section{Photonic memory}
\label{sec:mem}

\begin{figure}[t]
\centerline{\includegraphics[width=8.5cm]{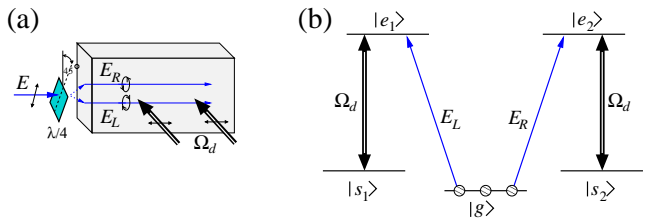}}
\caption{Reversible memory for the photon-polarization qubit.
(a) After the basis transformation $\ket{V} \to \ket{R}$,
$\ket{H} \to \ket{L}$ on a $45^{\circ}$ oriented $\la/4$ plate,
the photon enters the atomic medium serving as a memory device. 
(b) Level scheme of M-atoms interacting with the $E_{L,R}$ 
components of single-photon pulse and the driving field 
with Rabi frequency $\Om_d$.}
\label{fig:phmem}
\end{figure}

With straightforward modifications, the technique described above can be 
used to realize a reversible memory device for the photon-polarization qubit.
To that end, after passing through a $\la/4$ plate oriented at $45^{\circ}$, 
the vertically- and horizontally-polarized components of the single-photon 
pulse are converted into the circularly right- and left-polarized ones 
according to $\ket{V} \to \ket{R} = \frac{1}{\sqrt{2}} (\ket{V} + i \ket{H})$
and $\ket{H} \to \ket{L} = \frac{1}{\sqrt{2}} (\ket{V} - i \ket{H})$.
The pulse is then sent to the atomic medium with M-level configuration,
as shown in figure~\ref{fig:phmem}. All the atoms are initially prepared 
in the ground state $\ket{g}$, the $E_{L,R}$ components of the field 
interact with the atoms on the corresponding transitions 
$\ket{g} \to \ket{e_{1,2}}$, while the excited states $\ket{e_{1,2}}$ 
are coupled to the metastable states $\ket{s_{1,2}}$ via the same driving 
field with Rabi frequency $\Om_d$. Once the pulse has fully accommodated in 
the medium, the driving field $\Om_d$ is adiabatically switched off. As a 
result, the photon wavepacket is stopped in the medium, and its state 
$\ket{\psi}$ is coherently mapped onto the collective atomic state 
according to \cite{LYF,MFCM}
\begin{equation}
\alpha \ket{L} + \beta \ket{R} \to 
\alpha \ket{s_1^{(1)}} + \beta \ket{s_2^{(1)}} .
\label{phstore}
\end{equation}
This collective state is stable against decoherence \cite{MFCM}, 
allowing for a long storage time of the qubit state in the atomic 
ensemble. At a later time, the photon can be released from the medium 
on demand by switching the driving field on, which results in the 
reversal of mapping~(\ref{phstore}).

\section{Deterministic source of single-photons}
\label{sec:sphs}

Generation of single-photons in a well defined spatiotemporal mode is a 
challenging task that is currently attracting much effort \cite{SPHS-NJP}. 
In a simplest setup, one typically employs spontaneous parametric down 
conversion \cite{SPDC} to generate a pair of polarization- and 
momentum-correlated photons. Then, conditional upon the outcome of 
measurement on one of the photons, the other photon is projected onto 
a well-defined polarization and momentum state. In more elaborate 
experiments, single emitters, such as quantum dots \cite{sphsQDs} or 
molecules \cite{sphsMol}, emit single photons at a time when optically 
pumped into an excited state. Recently, truly deterministic sources of 
single photons have been realized with single atoms in high-$Q$ optical 
cavities \cite{sphsCRAP}. In these experiments, single-photon wavepackets 
with precise propagation direction and well characterized timing and 
temporal shape were generated using the technique of intracavity 
stimulated Raman adiabatic passage (STIRAP). Notwithstanding these 
achievements, the cavity QED experiments in the strong coupling regime 
required by the intracavity STIRAP involve sophisticated experimental 
setup which is very difficult, if not impossible, to scale up to a 
large number of independent emitters operating in parallel. 

In this section, we describe a method for deterministic generation of 
single-photon pulses from coherently manipulated atomic ensembles. As 
discussed in section~\ref{sec:eit}, symmetric Raman (spin) excitations in 
optically thick atomic medium exhibit collectively enhanced coupling to 
light in the EIT regime; Once the single excitation state $\ket{s^{(1)}}$ 
is created in the medium, the application of resonant driving field $\Om_d$ 
on the transition $\ket{s} \to \ket{e}$ will stimulate the Raman transition 
$\ket{s} \to \ket{g}$ and produce a single-photon anti-Stokes pulse $E$, 
whose propagation direction and pulse-shape are completely determined by 
the driving-field parameters. The question is then: How can one produce, 
in a deterministic fashion, precisely one collective Raman excitation? 
In a number of recent theoretical and experimental studies, such excitations
were produced by the process of spontaneous Raman scattering. Namely, one 
applies a classical pump laser to the atomic transition $\ket{g} \to \ket{e}$ 
and detects the number of forward scattered Stokes photons \cite{sphsSPRS}. 
Since the emission of each such photon results in one atomic 
excitation $\ket{s}$ symmetrically distributed in the whole ensemble, 
the number of Stokes photons is uniquely correlated with the number 
of Raman excitations of the medium. However, due to the spontaneous 
nature of the scattering process, the production of collective single
excitation state $\ket{s^{(1)}}$ requires the postselection conditioned 
upon the measured number of Stokes photons, which makes this scheme 
essentially probabilistic

Below we will describe a scheme that is capable of producing the single 
collective Raman excitation at a time. It is based on the dipole-blockade 
technique proposed in \cite{dipblk}, which employs the exceptionally 
strong dipole-dipole interactions between pairs of Rydberg atoms. In a 
static electric field $E_{\rm st} \textbf{e}_z$, the linear Stark effect 
results in splitting of highly excited Rydberg states into a manifold 
of $2 n - 1$ states with energy levels 
$\hbar \De \nu_{nqm} = \frac{3}{2} n q e a_0 E_{\rm st}$, where 
$n$ is the principal quantum number, 
$q \equiv n_1 - n_2 = n -1 - |m|, n - 3 - |m|, \ldots ,- (n -1 - |m|)$,
and $m = n-1, n-2,\ldots ,-(n -1)$ are, respectively, parabolic and magnetic
quantum numbers, $e$ is the electron charge, and $a_0$ is the Bohr radius 
\cite{RydAtoms}. These Stark eigenstates possess large permanent dipole 
moments $\textbf{d}= \frac{3}{2} n q e a_0 \textbf{e}_z$. A pair of atoms
1 and 2 prepared in such Stark eigenstates $\ket{r}$ interact with each 
other via the dipole-dipole potential
\begin{equation}
V_{\rm DD} = \frac{\textbf{d}_1 \cdot \textbf{d}_2 
- 3 (\textbf{d}_1 \cdot \textbf{e}_{12}) (\textbf{d}_2 \cdot \textbf{e}_{12})}
{4 \pi \eps_0 R^3} , \label{DDpot}
\end{equation}
where $\textbf{R} = R \textbf{e}_{12}$ is the distance between the atoms.
The dipole-dipole interaction (\ref{DDpot}) results in an energy shift of
the pair of Rydberg atoms, as well as their coupling to the other Stark 
eigenstates within the same $n$ manifold, which in turn splits the energy 
levels $\ket{r}$.

\begin{figure}[t]
\centerline{\includegraphics[width=3.5cm]{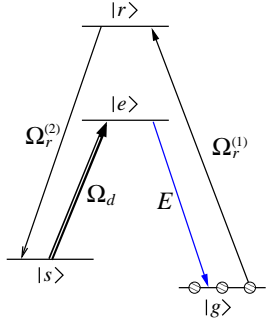}}
\caption{~Level scheme of atoms for deterministic generation of 
single-photons. Dipole-dipole interaction between pairs of atoms in
Rydberg states $\ket{r}$ facilitates the generation of single collective 
Raman excitation of the atomic ensemble at a time, via the sequential
application of the $\Om_r^{(1)}$ and $\Om_r^{(2)}$ pulses of (effective)
area $\pi$. This collective excitation is then adiabatically converted into 
a single-photon wavepacket by switching on the driving field $\Om_d$.}
\label{fig:sphs}
\end{figure}

Consider next a dense ensemble of double-$\La$ atoms shown in 
figure~\ref{fig:sphs}. A coherent laser field with Rabi frequency 
$\Om_r^{(1)} < \De \nu$ resonantly couples the initial atomic state 
$\ket{g}$ to the selected Stark eigenstate $\ket{r}$, while the second 
resonant field acts on the transition $\ket{r} \to \ket{s}$ with Rabi 
frequency $\Om_r^{(2)}$. When $\Om_d = E = 0$, one can disregard state 
$\ket{e}$, and the Hamiltonian takes the form
\begin{equation}
H = V_{\rm AF} + V_{\rm DD} ,
\end{equation}
with the atom-field and dipole-dipole interaction terms given, respectively, by
\begin{subequations}
\begin{eqnarray}
V_{\rm AF} &=& - \hbar \sum_j^N [\Om_r^{(1)} e^{ik_r^{(1)} z_j} \sih_{rg}^j 
+ \Om_r^{(2)} e^{ik_r^{(2)} z_j} \sih_{rs}^j 
\nonumber \\ & & \;\;\;\;\;\;\;\;\;\;\;\;
+ {\rm H. c.}], \\
V_{\rm DD} &=& \hbar \sum_{i > j}^N 
\De_{ij}(R) \ket{r_i \, r_j} \bra{r_i \, r_j} ,
\end{eqnarray}
\end{subequations}
where $\De_{ij}(R) = \bra{r_i \, r_j} V_{\rm DD} \ket{r_i \, r_j} 
\approx - n^4 e^2 a_0^2/(\pi \hbar \eps_0 R^3)$ is the dipole-dipole energy 
shift for a pair of atoms $i$ and $j$ separated by distance $R$. Suppose 
that initially all the atoms are in state $\ket{g}$, while the second 
laser is switched off, $\Om_r^{(2)} = 0$. Then the first laser field, 
coupled symmetrically to all the atoms, will induce the transition from 
the ground state $\ket{g_1,g_2,\ldots,g_N} \equiv \ket{s^{(0)}}$ to the 
collective state $\ket{r^{(1)}} \equiv \frac{1}{\sqrt{N}} \sum_j 
e^{i k_r^{(1)} z_j} \ket{g_1,\ldots,r_j, \ldots,g_N}$ representing
a symmetric single Rydberg excitation of the atomic ensemble. The
collective Rabi frequency on the transition $\ket{s^{(0)}} \to \ket{r^{(1)}}$
is $\sqrt{N} \Om_r^{(1)}$. Once an atom $i\,(\in \{1,\ldots,N \})$ is excited
to the state $\ket{r}$, the excitation of a second atom $j\,(\neq i)$ is 
constrained by the dipole-dipole interaction between the atoms: Provided 
$|\De_{ij}| > \Om_r^{(1)}, \ga_r$, where $\ga_r$ is the width of level
$\ket{r}$, the nonresonant transitions $\ket{r_i \, g_j} \to \ket{r_i \, r_j}$
resulting in two-atom excitations are suppressed. Hence, applying a laser 
pulse of area $\sqrt{N} \Om_r^{(1)} T = \pi /2$ (an effective $\pi$
pulse), one produces the single Rydberg excitation state $\ket{r^{(1)}}$.
At the end of the pulse, the probability of error due to populating the 
doubly-excited states $\ket{r_i \, r_j}$ is found by adding the
probabilities of all possible double-excitations,
\[
P_{\rm double} \sim \frac{1}{N}\sum_{i,j} \frac{|\Om_r^{(1)}|^2}{\De_{ij}^2}
\approx \frac{N |\Om_r^{(1)}|^2}{\bar{\De}^2} .
\] 
Thus $P_{\rm double} \ll 1$ when the collective Rabi frequency 
$\sqrt{N} \Om_r^{(1)}$ is small compared to the average dipole-dipole 
energy shift $\bar{\De}$. Another source of errors is the dephasing given 
by $P_{\rm deph} \leq \ga_r T \sim \ga_r/(\sqrt{N} \Om_r^{(1)})$, which is
typically very small for long-lived Rydberg states and $N \gg 1$. 
By the subsequent application of the second laser with the area 
$\Om_r^{(2)} T = \pi /2$ ($\pi$ pulse), the state $\ket{r^{(1)}}$ can be 
converted into the symmetric Raman excitation state 
$\ket{s^{(1)}} \equiv \frac{1}{\sqrt{N}} \sum_j 
e^{i (k_r^{(1)} - k_r^{(2)}) z_j} \ket{g_1,\ldots,s_j, \ldots,g_N}$,
which is precisely the state we need for the generation of single-photon
pulse, as described above.

To relate the foregoing discussion to a realistic experiment, let us 
assume cold alkali atoms (Rb) loaded into an elongated trap of length 
$L\simeq 10 \: \mu$m and cross-section $A \simeq 10 \: \mu {\rm m}^2$. 
The Stark eigenstates are resonantly selected from within the Rydberg states
with the effective principal quantum number $n\simeq 50$. The dipole-dipole 
energy shift is smallest for pairs of atoms located at the opposite ends of 
the trap, $\bar{\De} \gtrsim \De_{ij}(L) \sim 2\pi \times 20\:$MHz. 
For the density $\rho \simeq 10^{14} \: {\rm cm}^{-3}$, the trap contains 
$N \simeq 10^4$ atoms, and the (single atom) Rabi frequency should be chosen 
as $\Om_r^{(1)} \leq 2\pi \times 100\:$kHz. Then, for the preparation 
time $T \sim 0.1\:\mu {\rm s}$ of state $\ket{s^{(1)}}$, the achievable 
fidelity is $\gtrsim 98$\%. For these parameters, the optical depth
of the medium is large, $2 \ka_0 L \simeq 100$, which is necessary
for efficient generation of single-photon pulses by switching on the 
driving field $\Om_d$ and converting the atomic Raman excitation into
the photonic excitation, as discussed in section~\ref{sec:eit}.   

It should be mentioned that a related scheme for single photon generation 
employing the dipole blockade technique was proposed in \cite{sphsDDB}. 
There, however, a single atom at a time was transferred to an excited state,
from where it spontaneously decayed back to the ground state producing a 
single fluorescent photon in a well-defined direction.

\section{Photon-photon interaction}
\label{sec:xpm}

Conventional media are typically characterized by weak optical nonlinearities,
which are manifest only at high intensities of electromagnetic fields 
\cite{Boyd} and are vanishingly small for single- or few-photon fields. 
It was first pointed out in \cite{imam}, however, that the ultrahigh 
sensitivity of EIT dispersion to the two-photon Raman detuning $\de_R$ 
in the vicinity of absorption minimum, can be used to achieve giant Kerr 
nonlinearities between two weak optical fields interacting with 
four-level atoms in N-configuration of levels. As was shown in 
section~\ref{sec:eit}, a probe field with $\de_R=0$ propagating in 
the EIT medium experiences negligible absorption and phase-shift. 
When, however, a second weak (signal) field, dispersively coupling
state $\ket{s}$ to another excited state $\ket{f}$, is introduced 
in the medium, it causes a Stark shift of level $\ket{s}$, given
by $\De_{\rm St} = |\Om_s|^2/\De_s$, where $\Om_s$ is the Rabi frequency
and $\De_s > \Om_s,\Ga_f$ is the detuning of the signal field from the 
$\ket{s} \to \ket{f}$ resonance. Thus the EIT spectrum
is effectively shifted by the amount of $\De_{\rm St}$, which results 
in the conditional phase-shift of the probe field, $\phi(z) \simeq 
\ka_0 z \ga_{ge} \De_{\rm St}/|\Om_d|^2$. Notwithstanding this promising
sensitivity,  large conditional phase shift of one weak (single-photon)
pulse in the presence of another (also known as cross-phase modulation)
faces serious challenges in spatially uniform media. In order to
eliminate the two-photon absorption \cite{haryam} associated with Doppler
broadening $\de \om_{\rm D}$ of the atomic resonance $\ket{s} \to \ket{f}$,
one either has to work with cold atoms, of choose large detuning
$\De_s > \de \om_{\rm D}$, which limits the resulting cross-phase shift.
Another drawback of this scheme is the mismatch between the slow
group velocity of the probe pulse subject to EIT, and that of the 
nearly-free propagating signal pulse, which severely limits their 
effective interaction length and the maximal conditional phase-shift 
\cite{harhau}. This drawback may be remedied by using an equal mixture
of two isotopic species, interacting with two driving fields and an 
appropriate magnetic field, which would render the group velocities 
of the two pulses equal \cite{lukimam}. Other schemes to achieve the 
group velocity matching and strong cross-phase modulation were proposed
in \cite{DPGK,ottaviani,rebic}. Here we discuss an alternative, simple 
and robust approach \cite{DPYuM}, in which two weak (single-photon) 
fields, propagating through a medium of hot alkali atoms under the 
conditions of EIT, impress very large nonlinear phase-shift upon each 
other.  

\begin{figure}[t]
\centerline{\includegraphics[width=8.5cm]{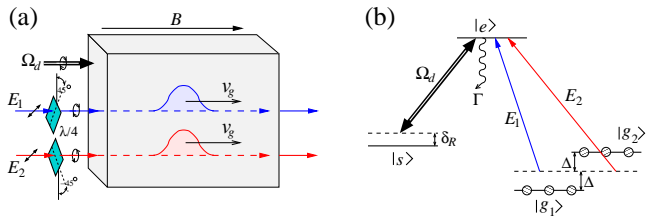}}
\caption{Cross-phase modulation of two single-photon pulses.
(a) Two horizontally polarized photons $E_1$ and $E_2$, after 
passing through the $\pm 45^{\circ}$ oriented $\la/4$ plates, 
are converted into the circularly left- and right-polarized photons, 
which are sent into the active medium. 
(b)~Level scheme of tripod atoms interacting with quantum fields 
$E_{1,2}$, strong cw driving field with Rabi frequency $\Om_d$ and 
weak magnetic field $B$ that removes the degeneracy of Zeeman 
sublevels $\ket{g_1}$ and $\ket{g_2}$.}
\label{fig:xphm}
\end{figure}

Consider a near-resonant interaction of two weak, orthogonally (circularly)
polarized optical fields $E_1$ and $E_2$ and a strong driving field with 
Rabi frequency $\Om_d$ with a medium of atoms having tripod configuration 
of levels, as shown in figure \ref{fig:xphm}. The medium is subject to a 
weak longitudinal magnetic field $B$ that removes the degeneracy of the 
ground state sublevels $\ket{g_1}$ and $\ket{g_2}$, whose Zeeman shift is 
given by $\hbar \De = \mu_{\rm B} m_F g_F B$, where $\mu_{\rm B}$ is the 
Bohr magneton, $g_F$ is the gyromagnetic factor and $m_F = \pm 1$ is the 
magnetic quantum number of the corresponding state. All the atoms are 
assumed to be optically pumped to the states $\ket{g_1}$ and $\ket{g_2}$,
which thus have the same incoherent populations $\expv{\sih_{g_1 g_1}} 
= \expv{\sih_{g_2 g_2}} = 1/2$. The weak fields $E_1$ and $E_2$, having 
the same carrier frequency $\om = \om^0_{eg}$ equal to the 
$\ket{g_{1,2}} \to \ket{e}$ resonance frequency for zero magnetic field, 
and wavevector $k$ parallel to the magnetic field direction, act on the 
atomic transitions $\ket{g_1} \to \ket{e}$ and  $\ket{g_2} \to \ket{e}$, 
with the detunings $\de_{1,2} = \mp \De - k v $, where $k v$ is the Doppler 
shift for the atoms having velocity $v$ along the propagation direction. 
In the collinear Doppler-free geometry shown in figure~\ref{fig:xphm}(a), 
the circularly polarized driving field couples level $\ket{e}$ to a 
single magnetic sublevel $\ket{s}$, whose Zeeman shift 
$\hbar \De^{\prime} = \mu_{\rm B} m_{F^{\prime}} g_{F^{\prime}}B$
is incorporated in the driving field detuning, $\de_d = \om_d-\om_{es}^0
+ \De^{\prime} - k_d v = \De_d - k_d v$.

Assuming, as before, the weak-field limit and adiabatically eliminating 
the atomic degrees of freedom, the equations of motion for the electric
field operators $\Eh_{1,2}$ corresponding to the quantum fields $E_{1,2}$ 
are obtained as \cite{DPYuM}
\begin{subequations}
\label{E12}
\begin{eqnarray}
\left( \ddz + \frac{1}{v_g} \ddt \right) \Eh_1 &=& 
- \ka_1 \Eh_1
-i (\De + \De_d) (s_1 - \eta_1 \Ih_2) \Eh_1 
\nonumber \\ & & \;\;\;\;
+ \Fch_1 , \\
\left( \ddz + \frac{1}{v_g} \ddt \right) \Eh_2 &=& 
- \ka_2 \Eh_2
+ i (\De - \De_d) (s_2 - \eta_2 \Ih_1) \Eh_2 
\nonumber \\ & & \;\;\;\;
+ \Fch_2 , 
\end{eqnarray}
\end{subequations}
where $v_g = c \cos^2 \theta \ll c$ is the group velocity, with the mixing 
angle $\theta$ defined as $\tan^2 \theta = g^2 N/(2|\Om_d|^2)$ (the factor 
1/2 corresponds to the initial population of levels $\ket{g_{1,2}}$), 
$\ka_{1,2} = \tan^2 \theta /c[\ga_R + \ga_{ge} (\De \pm \De_d)^2/|\Om_d|^2]$
and $s_{1,2} = \tan^2 \theta / c [1 + \De (\De \pm \De_d)/|\Om_d|^2]$ 
are, respectively, the linear absorption and phase modulation coefficients,
$\eta_{1,2} = g^2 2 \De \tan^2 \theta/[c |\Om_d|^2 (2\De \mp i \ga_R)]$
are the cross-coupling coefficients, and $\Ih_{j} \equiv \Eh_j^{\dagger} \Eh_j$
is the dimensionless intensity (photon-number) operator for the $j$th field.
In deriving equations~(\ref{E12}), the EIT conditions 
$|\Om_d|^2 \gg (\De +k \bar{v}) (\De \pm \De_d),\ga_R (\ga_{ge} + k \bar{v})$,
where $\bar{v}$ is the mean thermal atomic velocity, were assumed satisfied.
Note that if states $\ket{g_{1,2}}$ and $\ket{s}$ belong to 
different hyperfine components of a common ground state, the frequencies 
$\om$ and $\om_d$ of the optical fields differ from each other by at most
a few GHz, and the difference $(k-k_d) v$ in the Doppler shifts of the 
atomic resonances $\ket{g_{1,2}} \to \ket{e}$ and $\ket{s} \to \ket{e}$ 
is negligible. 

In what follows, we discuss the relatively simple case of small 
magnetic field, such that $\ga_R \ll \De,\De^{\prime} \ll \De_d$ 
and $\De_d=\om_d-\om_{es}^{0}$. When absorption is negligible, 
$\ka_{1,2} z \ll 1$, $z \in \{0,L \}$, which requires that 
$v_g / \ga_R \gg L$ and $\De_d^2 < \ga_R |\Om_d|^2/\ga_{ge}$, 
the solution of equations~(\ref{E12}) is
\begin{subequations}
\label{qE12slv}
\begin{eqnarray}
\Eh_1(z,t) &=& \Eh_1(0,\tau) 
\exp [ i \eta \De_d  \Eh_2^{\dagger}(0,\tau) \Eh_2(0,\tau) z ] , \\
\Eh_2(z,t) &=& \Eh_1(0,\tau) 
\exp [ i \eta \De_d  \Eh_1^{\dagger}(0,\tau) 
\Eh_1(0,\tau) z ] , 
\end{eqnarray}
\end{subequations}
where the cross-phase modulation coefficient is given by 
$\eta \simeq g^2/(v_g |\Om_d|^2)$, while the linear phase-modulation is 
incorporated into the field operators via the unitary transformations
$\Eh_{1,2}(z,t) \to \Eh_{1,2}(z,t) e^{i  \De_d z/v_g}$. The multimode 
field operators $\Eh_j(z,t) = \sum_{q} a_{j}^{q}(t) e^{i q z}$ ($j=1,2$),
with quantization bandwidth $\de q \leq \de \om_{\rm tw} /c$ 
($q \in \{-\de q/2, \de q/2\}$) restricted by the width of the EIT window
$\de \om_{\rm tw}$ \cite{lukimam}, have the following equal-time commutation 
relations
\[
[\Eh_{i}(z),\Eh_{j}^{\dagger}(\zp)]= 
\de_{ij} \frac{L \de q}{2 \pi} {\rm sinc}\left[ \de q (z-\zp)/2 \right] ,
\]
where ${\rm sinc}(x) = \sin(x)/x$.  

Consider the input state $\ket{\Phi_{\rm in}} = \ket{1_1} \otimes \ket{1_2}$,
consisting of two single photon wavepackets
\[
\ket{1_{j}} = \sum_{q} \xi_{j}^q a_{j}^{q \dagger} \ket{0} 
= \int d z f_j(z)\Eh_{j}^{\dagger}(z) \ket{0} \;\;\;  
(j = 1,2) ,
\] 
whose spatial envelopes $f_{j}(z) = \bra{0} \Eh_{j}(z,0) \ket{1_j}$ 
are initially (at $t=0$) localized around $z=0$.
The state of the system at any time can be represented as
\begin{equation}
\ket{\Phi(t)} = \sum_{q,q^{\prime}} \xi_{12}^{qq^{\prime}}(t)
\ket{1_1^q} \ket{1_2^{q^{\prime}}} , \label{st}
\end{equation}
from where it is apparent that $\xi_{12}^{qq^{\prime}}(0) = 
\xi_{1}^q \xi_{2}^{q^{\prime}}$. Since for the photon-number states the 
expectation values of the field operators vanish, all the information 
about the state of the system is contained in the intensities of the 
corresponding fields 
\begin{equation}
\expv{\Ih_j (z,t)} = \bra{\Phi_{\rm in}} \Eh_{j}^{\dagger} (z,t) 
\Eh_j (z,t)\ket{\Phi_{\rm in}}  , \label{evI}
\end{equation}
and their ``two-photon wavefunction'' \cite{ScZub,lukimam}
\begin{equation}
\Psi_{ij}(z,t;\zp,t^{\prime}) = 
\bra{0} \Eh_j(\zp,t^{\prime}) \Eh_i(z,t) \ket{\Phi_{\rm in}} .
\label{tphwf}
\end{equation}
The physical meaning of $\Psi_{ij}$ is a two-photon detection amplitude,
through which one can express the second-order correlation function
$G^{(2)}_{ij} = \Psi_{ij}^{*} \Psi_{ij}$ \cite{ScZub}. The knowledge 
of the two-photon wavefunction allows one to calculate the amplitudes 
$\xi_{12}^{qq^{\prime}}$ of state vector (\ref{st}) via the two 
dimensional Fourier transform of $\Psi_{ij}$ at $t = t^{\prime}$:
\begin{equation}
\xi_{ij}^{qq^{\prime}}(t) = \frac{1}{L^2} \int \!\!\! \int dz d \zp 
\Psi_{ij}(z,\zp,t) e^{-iqz}e^{-iq^{\prime} \zp } \label{ftrns}.  
\end{equation}

Substituting the operator solutions (\ref{qE12slv}) into  (\ref{evI}), for 
the expectation values of the intensities one finds 
\begin{equation}
\expv{\Ih_j (z,t)} = |f_j(-c \tau)|^2 .
\end{equation}
For $0 \leq z < L$ the retarded time is $\tau = t - z/v_g$, and therefore 
$\expv{\Ih_j (z,t)} = |f_j(z c /v_g - ct)|^2$, while outside the medium, 
at $z \geq  L$ and accordingly $\tau = t - L/v_g -(z-L)/c$, we have 
$\expv{\Ih_j (z,t)} = |f_j(z + L(c/v_g -1) -ct)|^2$. On the other hand,
after the interaction at $z,\zp \geq L$, the equal-time ($t = t^{\prime}$)
two-photon wavefunction reads \cite{DPYuM}
\begin{eqnarray}
\Psi_{ij}(z,\zp ,t) &=& f_i[z +L(c/v_g -1) -c t] 
\nonumber  \\ & & \times 
f_j[\zp + L(c/v_g -1) -c t]
\nonumber  \\ & & \times 
\left\{ 1 + 
\frac{f_j[z+ L(c/v_g -1) -c t]}{f_j[ \zp + L(c/v_g -1) -c t]} \:
\right. \nonumber  \\ & & \;\;\;\;\;\; \left. \times 
{\rm sinc}\left[ \frac{\de q}{2} (\zp - z) \right]
\left(e^{i \phi} -1 \right) \right\} , \;\;\;\; \label{et-tphwf}
\end{eqnarray}
where $\phi = \eta \De_d L^2 \de q /(2 \pi)$. For large enough spatial 
separation between the two photons, such that $|\zp - z| > \de q^{-1}$ 
and therefore ${\rm sinc} [ \de q (\zp - z)/2] \simeq 0$, 
equation~(\ref{et-tphwf}) yields
\[
\Psi_{ij}(z,\zp ,t) \simeq  
f_i[z +L(c/v_g -1) -c t] \: f_j[\zp + L(c/v_g -1) -c t] ,
\]
which indicates that no nonlinear interaction takes place between 
the photons, which emerge from the medium unchanged. This is due 
to the local character of the interaction described by the 
${\rm sinc}$ function. In the opposite limit of $|\zp - z| \ll \de q^{-1}$ 
and therefore ${\rm sinc} [ \de q (\zp - z)/2] \simeq 1$, for two 
narrow-band (Fourier limited) pulses with the duration 
$T \gg |\zp - z|/c$, one has $f_j(z)/f_j(\zp ) \simeq 1$, and 
equation~(\ref{et-tphwf}) results in 
\begin{eqnarray*}
\Psi_{ij}(z,\zp ,t) &\simeq & e^{i \phi}
f_i[z +L(c/v_g -1) -c t] 
\nonumber  \\ & & \;\;\;\; \times
f_j[\zp + L(c/v_g -1) -c t] .
\end{eqnarray*}
Thus, after the interaction, a pair of single photons acquires conditional
phase shift $\phi$, which can exceed $\pi$ when 
$(\de q L/2 \pi)^2 > (v_g/c) \, (|\Om_d|^2/g^2)$.
To see this more clearly, we use equation~(\ref{ftrns}) to calculate the
amplitudes of the state vector $\ket{\Phi(t)}$:
\begin{equation}
\xi_{ij}^{q q^{\prime}}(t) = e^{i \phi} \xi_{ij}^{q q^{\prime}}(0)  
\exp \{i (q + q^{\prime}) [L(c/v_g -1) -ct ] \} . \label{res-ftrns}
\end{equation}
At the exit from the medium, at time $t \simeq L/v_g$, the second exponent 
in equation~(\ref{res-ftrns}) can be neglected for all $q,q^{\prime}$ and 
the state of the system is given by 
\begin{equation}
\ket{\Phi(L/v_g)} = e^{i \phi}\ket{\Phi_{\rm in}} .
\end{equation} 
When $\phi = \pi$, the output state of the two photons is 
\begin{equation}
\ket{\Phi_{\rm out}} = - \ket{\Phi_{\rm in}} . \label{pi_phsh}
\end{equation} 
Utilizing the scheme of figure~\ref{fig:UWrztn}(b), one can then realize the 
transformation corresponding to the \textsc{cphase} logic gate between two 
photons representing qubits.

Before closing this section, we note that several important relevant issues,
such as the spectral broadening of interacting pulses and the necessity for 
their tight focusing over considerable interaction lengths, were addressed 
in a number of recent studies \cite{IFGKDP,MMMF,IFGKDPMF}.

\section{Single-photon detection}
\label{sec:sphd}

To complete the proposal for optical quantum computer, we need to 
discuss a measurement scheme capable of reliably detecting the 
polarization states of single photons. When the photonic qubit 
$\ket{\psi} = \alpha \ket{V} + \beta \ket{H}$ passes though a 
polarizing beam-splitter, its vertically and horizontally polarized 
components are sent into two different spatial modes -- photonic channels.
Placing efficient single-photon detectors at each channel would therefore 
accomplish the projective measurement of the qubits in the computational 
basis. The remaining question then is the practical realization of sensitive 
photodetectors. Avalanche photodetectors with high quantum efficiencies 
are possible candidates for the reliable measurement \cite{avlphd}. Let us,
however, outline an alternative scheme \cite{photdet}, based on stopping of
light in EIT media, whose potential efficiency is unmatched by the 
state-of-the-art photodetectors.

\begin{figure}[t]
\centerline{\includegraphics[width=4cm]{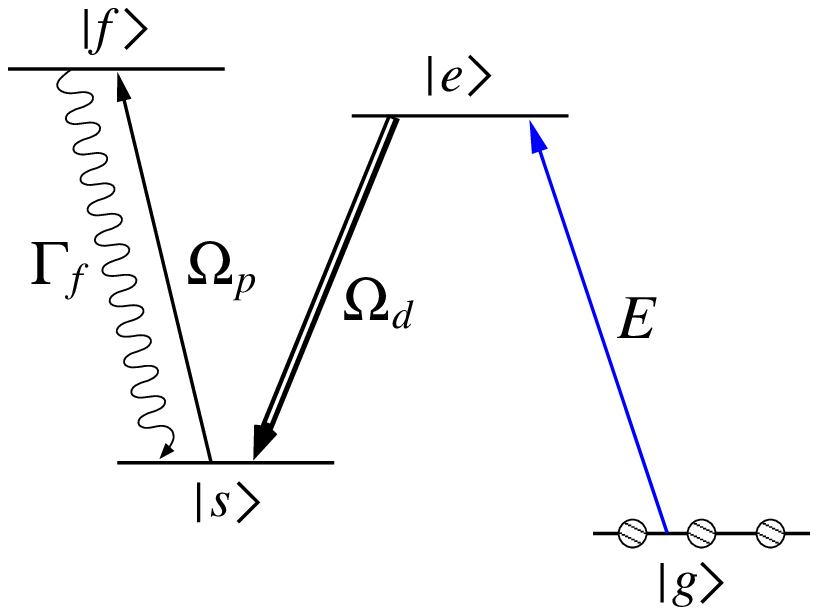}}
\caption{~Level scheme of atoms for efficient photon detection. 
Adiabatically switching off the driving field, $\Om_d \to 0$, 
results in the conversion of photonic excitation $E$ into the 
atomic Raman excitation $\ket{g} \to \ket{s}$. The latter is 
detected using the pump field acting on the cycling transition 
$\ket{s} \lra \ket{f}$ with Rabi frequency $\Om_p$ and collecting
the fluorescent photons.}
\label{fig:phdet}
\end{figure}

Consider an optically dense medium of four-level atoms with N-configuration
of levels, as shown in figure~\ref{fig:phdet}. Initially, all the atoms
are in the ground state $\ket{g}$. Under the EIT conditions discussed 
in section~\ref{sec:eit}, a single-photon pulse entering the medium
can be fully stopped by adiabatically switching off the driving field 
Rabi frequency $\Om_d$. As a result, the atomic ensemble is transferred
into the symmetric state $\ket{s^{(1)}}$ with single Raman excitation,
i.e., an atom in state $\ket{s}$. Next, to detect the atom in this state, 
one can use the electron shelving or quantum jump technique \cite{QJumps}.
To that end, one applies a strong resonant pumping laser acting on the
cycling transition $\ket{s} \lra \ket{f}$ with Rabi frequency $\Om_p$
and collects the fluorescent photons emitted by the atoms with the 
rate $R_f = \Ga_f |\Om_p|^2/(2 |\Om_p|^2 + \ga_{sf}^2)$, where 
$\Ga_f$ is the spontaneous decay rate of state $\ket{f}$. In the limit
of strong pump $\Om_p \gg \ga_{sf}$, transition $\ket{s} \to \ket{f}$
saturates and $R_f \sim \Ga_f/2$. 

In alkali atoms, the cycling transition with circularly polarized pump 
laser can be established between the ground and excited state sublevels 
$\ket{s} = \ket{F=2,m_F = 2}$ and $\ket{f} = \ket{F=3,m_F = 3}$.
To estimate the reliability of the measurement, assuming unit 
probability of photon trapping in EIT medium, let as suppose that
a photodetector with efficiency $\eta \ll 1$ is collecting the 
fluorescent signal $S_f = \eta R_f t$ during time $t$. This
time is limited by the lifetime of state $\ket{s}$, $\Ga_s^{-1}$,
which is related to the Raman coherence relaxation rate by
$\ga_R \geq \Ga_s/2$. A reliable measurement requires 
$S_f = \frac{1}{2}\eta (\Ga_f/\Ga_s) \geq 1$. Typically, in atomic 
vapors the ratio $\Ga_f/\Ga_s \sim 10^4$, therefore the signal $S_f$
is very strong even for tiny efficiencies $\eta \gtrsim 10^{-3}$.
Thus the described scheme offers great sensitivity in single- of
few-photon detection.

\section{Conclusions}
\label{sec:sum}

In this paper, we have described a proposal for all-optical deterministic 
quantum computation with photon-polarization qubits. The schemes for
deterministic generation of single-photon wavepackets, their storage, 
manipulation, entanglement and reliable measurement were discussed. 
All these schemes are based on the coherent manipulation of macroscopic 
atomic ensembles in the regime of electromagnetically induced transparency,
whose concise yet detailed description was presented for the sake of 
clarity and accessibility of presentation.

We have outlined the principal setup of the quantum computer and its 
building blocks, leaving out detailed studies of several important issues 
pertaining to the decoherence mechanisms and fidelity of the computer's 
constituent parts and their optimization, which will be addressed in 
subsequent publications. Certainly, the scheme described above is open 
to modifications and improvements, while some of its ingredients are 
still in the conceptual stage and have not yet been realized experimentally.
It seems therefore conceivable that at least in the short term, an 
optimized combination of the two approaches, linear optics probabilistic 
\cite{linopt,sphsSPRS} and nonlinear deterministic discussed here, would 
constitute the most realistic way towards the all-optical quantum computation.

\begin{acknowledgments}
I would like to thank  M.~Fleischhauer, I.~Friedler, G.~Kurizki, and 
Yu.P.~Malakyan for many useful discussions and fruitful collaboration.
\end{acknowledgments}

\end{document}